\documentclass{article} \usepackage[utf8]{inputenc} \usepackage{amsmath}
\usepackage{amsfonts} \usepackage{amssymb} \usepackage{amsthm}
\usepackage{stmaryrd} \usepackage{tikz}

\newtheorem{theorem}{Theorem}[section] \newtheorem{lemma}[theorem]{Lemma}

\definecolor{dgreen}{rgb}{0,0.6,0} \definecolor{dred}{rgb}{0.6,0,0}
\definecolor{dblue}{rgb}{0,0,0.6} \definecolor{lgray}{rgb}{0.8,0.8,0.8}
\definecolor{lblue}{rgb}{0.6,0.6,1} \definecolor{lgreen}{rgb}{0.6,1,0.6}
\definecolor{lred}{rgb}{1,0.6,0.6}

 \newcommand{\Vertex}[1]{\node[circle,inner sep = 0pt,minimum size
    =5pt,fill = #1]}

\newcommand{\background}[1]{ \draw (0,0) circle (1); \draw[black,dashed,thin]
  (270:1.3) -- (90:1.3); \draw (0.4,1.2) node {R}; \draw (-0.4,1.2) node {L};
  \Vertex{black} (u1) at (306:1) {}; \draw (306:1.1) node {$u_1$};
  \Vertex{black} (u2) at (342:1) {}; \draw (342:1.1) node {$u_2$};
  \Vertex{black} (u3) at (18:1) {}; \draw (18:1.1) node {$u_3$};
  \Vertex{black} (u4) at (54:1) {}; \draw (54:1.1) node {$u_4$};
  \Vertex{black} (u5) at (126:1) {}; \draw (126:1.1) node {$u_5$};
  \Vertex{black} (u6) at (162:1) {}; \draw (162:1.1) node {$u_6$};
  \Vertex{black} (u7) at (198:1) {}; \draw (198:1.1) node {$u_7$};
  \Vertex{black} (u8) at (234:1) {}; \draw (234:1.1) node {$u_8$};
  \Vertex{black} (a) at (282:1) {}; \Vertex{black} (b) at (318:1) {};
  \Vertex{black} (c) at (330:1) {}; \Vertex{black} (d) at (18:1) {};
  \Vertex{black} (g) at (180:1) {}; \Vertex{black} (h) at (258:1) {}; }

\title{Congestion in planar graphs with demands on faces} \author{Guyslain
  Naves\footnote{Department of Mathematics and Statistics, McGill University,
    {\texttt naves@math.mcgill.ca}},
  Christophe Weibel\footnote{Department of Computer Science, Dartmouth
    College, {\texttt weibel@math.mcgill.ca}}} \date{August 20, 2010}

\begin{document}

\maketitle

We show the following theorem:

\begin{theorem}\label{th:main}
The congestion in $(G,H,r,c)$ is at most $2 \lceil \log_2 k\rceil + 2$ when
$G$ is an embedded planar graph, each demand $h \in H$ lies on a face of $G$,
and there are at most $k$ terminals in each face of $G$.
\end{theorem}

We will use the celebrated theorem of Seymour:
\begin{theorem}\label{th:seymour}~\cite{Seymour81}
Let $(G,H,r,c)$ be an instance of the multiflow problem such that $G+H$ has no
$K_5$-minor. Then the cut condition is equivalent to the existence of a
half-integer multiflow. If $G+H$ is Eulerian, the cut condition is equivalent
to the existence of an integer multiflow.
\end{theorem}
Note that by Kuratowski's theorem, planar graphs have no $K_5$ minor.

\section{Proof}

Let $G$ be a planar graph. Without loss of generality, we suppose that $G$ is
$2$-connected. This means that the boundaries of its faces are circuits. For
any instance $(G,H,r,c)$, we define $r(e) = 0$ for every edge $e \notin E(H)$.

We say that two demand edges $s_1t_1$, $s_2t_2$ are crossed if they both lie
on the same face of $G$ and $s_1$, $s_2$, $t_1$, $t_2$ appears in that order
around the boundary of the face. Let $m$ be the minimum of $r(s_1t_1)$ and
$r(s_2t_2)$, we call \emph{uncrossing $(G,H)$ by $s_1t_1, s_2t_2$} and denote
$(G,H,r,c) \oplus (s_1t_1,s_2t_2)$ the instance $(G,H',r',c)$ where:
\begin{itemize}
\item[-] $r'(s_1t_1) = r(s_1t_1) - m$ and $r'(s_2t_2) = r(s_2t_2) - m$,
\item[-] $r'(s_1s_2) = r(s_1s_2) + m$ and $r'(t_1t_2) = r(t_1t_2) + m$,
\item[-] $r'(e) = r(e)$ for every other edge $e$,
\item[-] $H' = \{uv~:~r'(uv) > 0\}$.
\end{itemize}

\begin{lemma}\label{lemma:uncrossing}
Let $G$ be an embedded planar graph, $H$ a demand graph for $G$, and $s_1t_1$,
$s_2t_2$ two demands of $H$ lying on the same face of $G$. If $(G,H,r,c)$
satisfies the cut condition, so does $(G,H,r,c) \oplus (s_1t_1,s_2t_2)$.
\end{lemma}

\begin{proof}
It follows from the fact that the cut condition is satisfied iff it is
satisfied for central cuts only (\emph{i.e.} cuts $C = \delta(X)$ where $X$
and its complement are both connected in $G$). But the intersection of a
central cut and the boundary of a face is a path. From
this, the proposition can be easily checked.
\end{proof}

As a consequence, for any set of disjoint crossed demand edges, the cut
condition for $(G,H)$ implies the cut condition for the uncrossing of $(G,H)$
by these crossed demand edges.

From now on, we suppose the cut condition is satisfied by $(G,H,r,c)$. Let $F$
be any face of $G$ that contains some demand edges $H_F$. If $G + H_f$ is
planar, then by doubling $r$ and $c$ and applying the Eulerian part of
Theorem~\ref{th:seymour}, $G + H_f$ has congestion two. Note that actually,
for any number of faces $H_{F_1}, \ldots, H_{F_i}$, if $G + H_{F_1} + \ldots +
H_{F_i}$ is planar, its congestion is $2$. For convenience, we will only look
at one face at a time, but all the arguments can (and must) be applied
simultaneously on all the faces. We use this principle to decrease by half the
maximum number of terminals on one face of the demand graph. Note that when a
face $F$ has a single or no demand, $G + H_F$ is obviously planar.

Let $F$ be a face with a least two demands. For convenience, we only consider
the vertices of the boundary of $F$ that are terminals of the demand lying in
$F$, call them $u_1, u_2, \ldots,u_m$ (in the order of appearance on the
boundary), where $m = |V(H_F)|$. Let $k = \lfloor \frac{m}{2} \rfloor$. A
demand edge is \emph{bilateral} if one if its extremity is in $R =
\{u_1,\ldots,u_k\}$ and the other is in $L = \{u_{k+1},\ldots,u_m\}$.  We want
to route all the bilateral demands with a congestion of $2$. Then we would add
an edge of capacity $0$ between $u_k$ and $u_m$, completing the
proof. Actually, we will not solve these demands, but we will uncross all of
them in such a way that the new demands will have their two extremities both
in $L$ or both in $R$.

We define iteratively crossed pairs of bilateral edges of $H_F$. Let $i$ be
the minimum index such that there is a bilateral edge $u_iu_j$ in $H_F$, with
$j$ maximal. Let $j'$ be the maximum index such that there is a bilateral edge
$u_{i'}u_{j'}$ in $H_f$, with $i$ minimal. Note that $i$ exists iff $j'$
exists. Let $m = \min \{r(u_iu_j),r(u_{i'}u_{j'})\}$. We distinguish two
cases:
\begin{itemize}
\item[-] either $i=i'$ and $j=j'$, then we mark $u_iu_j$ in white,
\item[-] or we select the crossed edges $u_iu_j$ and $u_{i'}u_{j'}$, and mark
  $u_iu_{j'}$ in white.
\end{itemize}
In both cases, we decrease the requests on the edges $u_iv_j$ and $v_k, u_l$
by $m$ and remove the demand edges with capacity $0$. We repeat this procedure
until there is no more edges between $u_1,\ldots,u_m$ and $v_1,\ldots,v_m$.

\begin{figure}
\begin{center}
\begin{tikzpicture}[x=2.7cm,y=2.7cm]
\background{} \draw[thick] (u1) -- (u7) node[midway,below,blue] {$3$};
\draw[thick] (u2) -- (u7) node[below,pos=0.4,blue] {$1$}; \draw[thick] (u3) --
(u5) node[above,pos=0.8,blue] {$2$}; \draw[thick] (u4) -- (u8)
node[above,pos=0.6,blue] {$6$}; \draw[thick] (u4) -- (u6)
node[below,pos=0.7,blue] {$5$}; \draw (u3) .. controls (36:0.8) .. (u4)
node[sloped,above,midway,dgreen] {$3$}; \draw (a) .. controls (306:0.6) .. (c)
node[sloped,below,midway,dgreen] {$4$}; \draw (b) .. controls (348:0.6) .. (d)
node[sloped,below,midway,dgreen] {$2$}; \draw (g) .. controls (219:0.5) .. (h)
node[sloped,above,pos=0.3,dgreen] {$1$};
\end{tikzpicture}
\end{center}
\caption{The original face. The capacities of the bilateral demands are in
  blue, other demands are in green.}
\label{fig:fig1}
\end{figure}
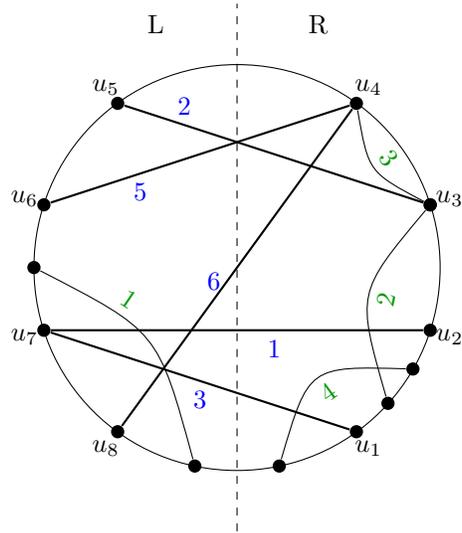

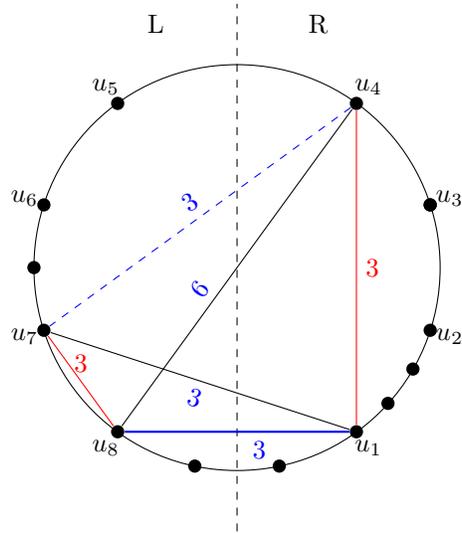
\begin{figure}
\begin{center}
\begin{tikzpicture}[x=2.7cm,y=2.7cm]
\background{} \draw (u1) -- (u7) node[sloped,midway,below,blue] {$3$}; \draw
(u4) -- (u8) node[sloped,above,pos=0.6,blue] {$6$}; \draw[thick,blue] (u1) --
(u8) node[sloped,below,pos=0.4,blue] {$3$}; \draw[dashed,blue] (u4) -- (u7)
node[sloped,above,midway,blue] {$3$}; \draw[red] (u1) -- (u4)
node[right,midway,red] {$3$}; \draw[red] (u8) -- (u7) node[above,midway,red]
{$3$};
\end{tikzpicture}
\end{center}
\caption{First iteration, $i=1$, $j=7$, $j'=8$ and $i'=4$. The minimum demand
  here is $3$, we decrease the capacities of these two bilateral demands by
  $3$. The blue continuous egde is marked white. The red edges are the result
  of the uncrossing.}
\label{fig:fig2}
\end{figure}

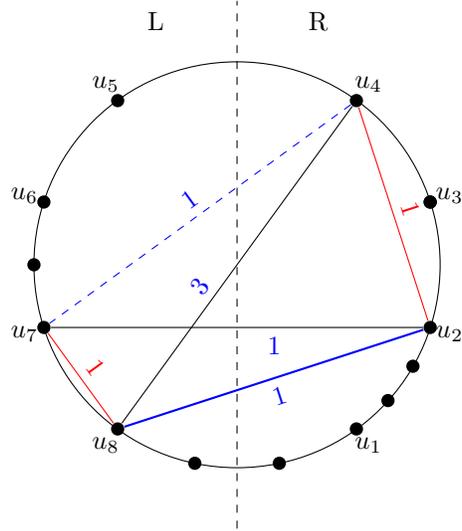
\begin{figure}
\begin{center}
\begin{tikzpicture}[x=2.7cm,y=2.7cm]
\background{} \draw (u2) -- (u7) node[sloped,pos=0.4,below,blue] {$1$}; \draw
(u4) -- (u8) node[sloped,above,pos=0.6,blue] {$3$}; \draw[thick,blue] (u2) --
(u8) node[sloped,below,midway,blue] {$1$}; \draw[dashed,blue] (u4) -- (u7)
node[sloped,above,midway,blue] {$1$}; \draw[red] (u2) -- (u4)
node[sloped,above,midway,red] {$1$}; \draw[red] (u8) -- (u7)
node[sloped,above,midway,red] {$1$};
\end{tikzpicture}
\end{center}
\caption{Second iteration, between edges $u_2u_7$ and $u_4u_8$, with minimum
  capacity $1$. The edge $u_2u_8$ is marked in white.}
\label{fig:fig3}
\end{figure}

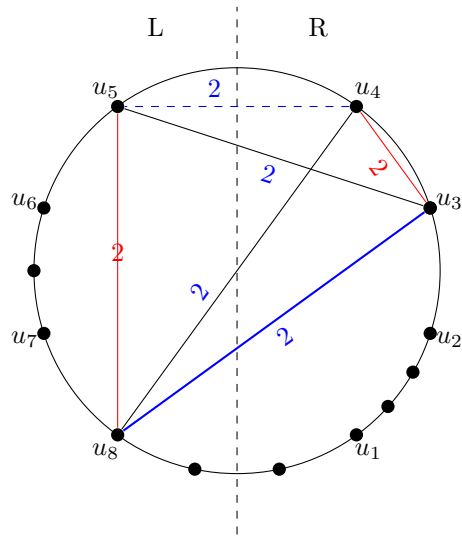
\begin{figure}
\begin{center}
\begin{tikzpicture}[x=2.7cm,y=2.7cm]
\background{} \draw (u3) -- (u5) node[sloped,midway,below,blue] {$2$}; \draw
(u4) -- (u8) node[sloped,above,pos=0.6,blue] {$2$}; \draw[thick,blue] (u3) --
(u8) node[sloped,below,midway,blue] {$2$}; \draw[dashed,blue] (u4) -- (u5)
node[sloped,above,pos=0.6,blue] {$2$}; \draw[red] (u3) -- (u4)
node[sloped,below,midway,red] {$2$}; \draw[red] (u8) -- (u5)
node[above,midway,red] {$2$};
\end{tikzpicture}
\end{center}
\caption{Third iteration, between edges $u_3u_5$ and $u_4u_8$. $u_3u_8$
  becomes white.}
\label{fig:fig4}
\end{figure}

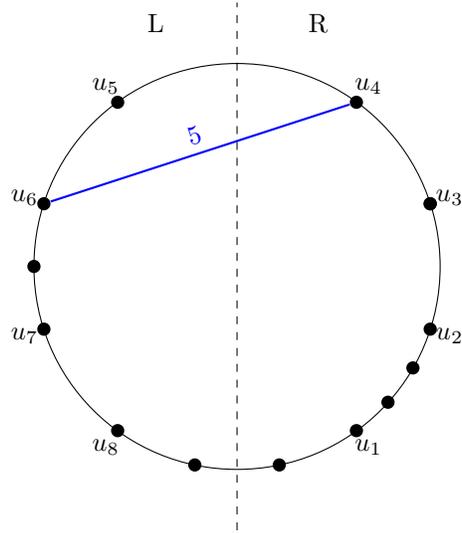
\begin{figure}
\begin{center}
\begin{tikzpicture}[x=2.7cm,y=2.7cm]
\background{} \draw[thick,blue] (u4) -- (u6) node[sloped,above,midway,blue]
           {$5$};
\end{tikzpicture}
\end{center}
\caption{Last iteration, this time $i = j'$ and $j = i'$. $u_4u_6$ is marked
  in white.}
\label{fig:fig5}
\end{figure}

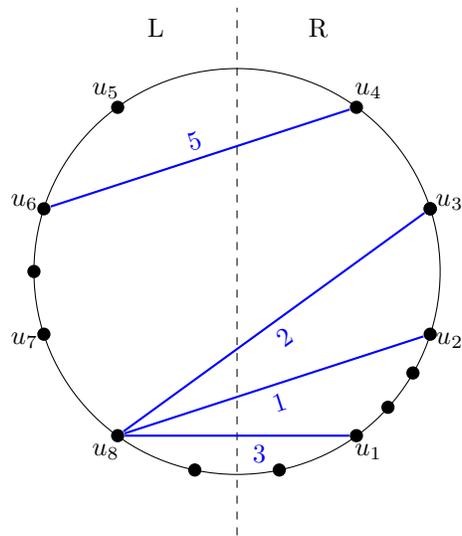
\begin{figure}
\begin{center}
\begin{tikzpicture}[x=2.7cm,y=2.7cm]
\background{} \draw[thick,blue] (u1) -- (u8) node[sloped,below,pos=0.4,blue]
           {$3$}; \draw[thick,blue] (u2) -- (u8)
           node[sloped,below,midway,blue] {$1$}; \draw[thick,blue] (u3) --
           (u8) node[sloped,below,midway,blue] {$2$}; \draw[thick,blue] (u4)
           -- (u6) node[sloped,above,midway,blue] {$5$};
\end{tikzpicture}
\end{center}
\caption{The white edges (in blue) are uncrossed. Their capacities are given
  by the uncrossing lemma, applied to the selected crossed pairs..}
\label{fig:fig6}
\end{figure}

\begin{figure}
\begin{center}
\begin{tikzpicture}[x=2.7cm,y=2.7cm]
\background{} \draw[red] (u1) .. controls (0:0.2) .. (u4)
node[left,midway,red] {$3$}; \draw[red] (u2) .. controls (18:0.4) .. (u4)
node[right,midway,red] {$1$}; \draw[red] (u5) .. controls (180:0.2) .. (u8)
node[right,midway,red] {$2$}; \draw[red] (u7) .. controls (216:0.8) .. (u8)
node[below,midway,sloped,red] {$4$}; \draw[red] (u3) .. controls (36:0.8)
.. (u4) node[sloped,above,midway,dgreen] {$3 {\color{red} + 2}$}; \draw (a)
.. controls (306:0.6) .. (c) node[sloped,below,midway,dgreen] {$4$}; \draw (b)
.. controls (348:0.6) .. (d) node[sloped,below,midway,dgreen] {$2$}; \draw (g)
.. controls (219:0.5) .. (h) node[sloped,above,pos=0.3,dgreen] {$1$};
\end{tikzpicture}
\end{center}
\caption{After uncrossing, there is no more bilateral edge.}
\label{fig:fig7}
\end{figure}
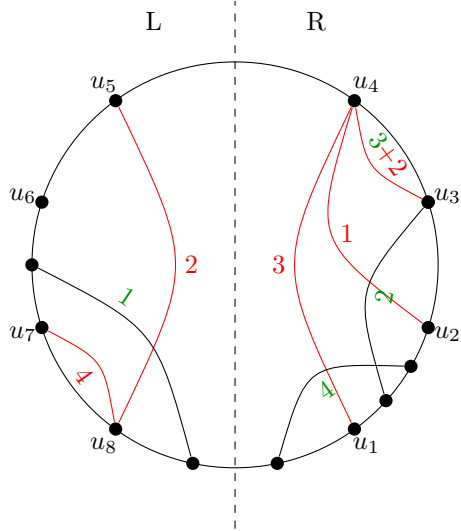

Thus, we have a set $S$ of selected crossed disjoint pairs of demands and a
set $W$ of white edges. By induction, it is easy to see that there are no two
crossed white edges. Moreover, by Lemma~\ref{lemma:uncrossing}, the two
following instances satisfies the cut condition:
\begin{itemize}
\item[$(i)$] $(G,H,r,c) \oplus \bigoplus_{(u_iu_j,u_{i'}u_{j'}) \in S}
  (u_iu_{j},u_{j'}u_{i'})$
\item[$(ii)$] $(G,H,r,c) \oplus \bigoplus_{(u_iu_j,u_{i'}u_{j'}) \in S}
  (u_iu_{j},u_{i'}u_{j'})$
\end{itemize}

By $(i)$, $(G,W)$ also satisfies the cut condition (by simply removing the
non-white demand edges). By Theorem~\ref{th:seymour}, $(2G, 2W)$ admits an
integer solution. From this solution, we only keep two paths for each unit of
capacity of the edge $u_iu_{j'}$, for each $(u_iu_j,u_{i'}u_{j'}) \in S$. For
all the edges $u_iv_j \in W \setminus E(S)$, we keep as many paths as the
capacity. This means that now we only have to find paths for each of the
demands $u_iu_{i'}$ and $u_ju_{j'}$ (and combine them with the two
$(u_i,u_{j'})$-paths), for each selected pair $(u_iu_j,u_{i'}u_{j'})$, plus
paths for all the non-bilateral demands. It corresponds to $(ii)$ without the
edges in $W \setminus E(S)$, thus it satifies the cut condition, and there is
no bilateral demand edge. By adding one supply edge with capacity $0$ between
$u_m$ and $u_k$ (it obviously does not violate the cut condition, nor does it
changes the feasibility of the instance), we obtain two new faces with at most
half the number of terminals of the original face.

By applying this procedure simultaneously (that is with only one invocation of
Theorem~\ref{th:seymour}) to every face, the maximal number of terminals in
one face is divided by two. Now, by induction, as each step uses $2G$, the
Theorem~\ref{th:main} is proved.

\section{Lower bound}

We now prove that one cannot largely improve our bound on congestion by simply
using Seymour's Theorem~\ref{th:seymour} as we did. More precisely, suppose we
apply Theorem~\ref{th:seymour} $c$ times to a face $F$ containing a set $T$ of
$n$ terminals. Without loss of generality, we prove the bound for the case
when $H_F$ is a matching. For each application, we get a solution to a planar
demand graph on $F$, with at most $2n$ arcs of demand. Then, at the end, we
have $2nc$ paths between the terminals on the boundary of $F$. We want to use
these paths to route the original demands $H_F$.

First, the number of possible planar demand graphs on $F$ with maximum degree
$2$ is equal to the number of noncrossing partitions of $T$. A
\emph{noncrossing partition} of a set $T = \{t_1,\ldots,t_n\}$ is a partition
without two parts $A$ and $B$, such that there are $i < j < k < l$ with
$t_i,t_k \in A$ and $t_j, t_l \in B$. The number of noncrossing partitions is
well-known to be the $n$th Catalan number $C_n =
\frac{1}{n+1}\binom{2n}{n}$~\cite{Kreweras72}. As we take $c$ of these graphs,
there is at most $C_n^c$ possible choices of $2nc$ paths by this method.

Then, let $\mathcal{P}$ be a set of $2nc$ paths on $n$ terminals, each
terminal having $2c$ paths ending at it. We want to glue together paths from
$\mathcal{P}$ in order to get a solution to our original problem. A part will
contain an ordering $P_1,\ldots,P_k$ of its paths, where $P_i$ is a
$(u_i,u_{i+1})$-path. Such a part satisfies the original demand edge
$(u_1,u_{k+1})$. Thus, we need to give an upper bound on the number of
partitions of $\mathcal{P}$ in consecutive sub-paths of a path. We can
represent $\mathcal{P}$ as a $2c$-regular graph $H'$ with $n$ vertices and
$2nc$ edges. We are looking for the number of partition of $H'$ into
paths. But a partition into paths can be encoded in the following way: for
each vertex $v$, give a perfect matching on $\delta(v)$. Two edges incident to
$v$ are matched if they are consecutive in one of the paths of the
partition. As this creates a partition into cycles, we also need to choose one
of the $2c$ incident edges to be the extremity of a path.

An upper bound on the number of partition can then be deduced from an upper
bound on the number of perfect matchings in the complete graph with $2c$
vertices, times $2c$. This last value is given by
\begin{equation}
m_c = \frac{(2c)!}{2^cc!}2c
\end{equation}
So given one of the $C_n^c$ possible choices of $c$ planar demand graphs, we
get an upper bound of $m_c^{2n}$ possible partitions into paths.  It proves
that the number of planar or non-planar demand graphs on $T$ that can be
solved by $c$ applications of Theorem~\ref{th:seymour} is at most $m_c^{2n}
C_n^c$. But the total number of possible demand graphs is
$\frac{(2n)!}{n!2^n}$, and the following analysis shows that we need $c =
\Omega\left(\frac{\log n}{\log\log n}\right)$.

We prove this by showing that if $c=\frac{\log n}{4\log\log n}-2$, $m_c^{2n}
C_n^c$ is asymptotically smaller than $\frac{(2n)!}{n!2^n} = (2n-1)!!$.
First, we have that
$$ m_c = \frac{(2c)!}{c!2^c}2c = (2c-1)!!2c \leq \frac{(2c)!!}{2}2c =
2^{c-1}c!2c \leq 2^c(c+1)! \leq 2^ce\left(\frac{c+2}{e}\right)^{c+2} =
$$
$$ \frac{e}{4}\left(\frac{2(c+2)}{e}\right)^{c+2} \leq
\left(\frac{2(c+2)}{e}\right)^{c+2}
$$ Considering $C_n$ is the number of correctly-matched parentheses, it is
trivial that $C_n \leq 2^{2n}$.  And so we can write
$$ m_c^{2n} C_n^c \leq \left(\frac{2(c+2)}{e}\right)^{(c+2)2n} 2^{2nc} \leq
\left(\frac{2(c+2)}{e}\right)^{(c+2)2n} 2^{(c+2)2n} =
$$
$$ \left(\frac{4(c+2)}{e}\right)^{(c+2)2n} \leq \frac{1}{e^n} 4(c+2)^{(c+2)2n}
$$ If we replace $(c+2)$ with $\frac{\log n}{4\log\log n}$, we get:
$$ m_c^{2n} C_n^c \leq \frac{1}{e^n} \left(\frac{\log n}{\log\log
  n}\right)^{\frac{2n\log n}{4\log\log n}} \leq \frac{1}{e^n} (\log
n)^{\frac{n\log n}{2\log\log n}} = \frac{1}{e^n} e^{\frac{n \log n \log\log
    n}{2\log\log n}} =
$$
$$ \frac{1}{e^n} e^{\frac{n\log n}{2}} = \frac{1}{e^n} n^{\frac{n}{2}} <
\left(\frac{n}{e}\right)^n < e \left(\frac{n}{e}\right)^n < n! < 2^{n-1}
(n-1)! = (2n-2)!! \leq (2n-1)!!
$$\bigskip

{\bfseries\noindent Acknowledgments}\\
The authors thank Chandra Chekuri and Bruce Shepherd for their useful remarks
regarding the lower bound of our method.

\bibliographystyle{abbrv}

\end{document}